# Enhanced third-harmonic generation induced by nonlinear field resonances in plasmonic-graphene metasurfaces

Yulian Liu[1], Shan Zhu[2], Qingjia Zhou[1], Yanyan Cao[1], Yangyang Fu[3, *], Lei Gao[1, *], Huanyang Chen[2] and Yadong Xu[1, *]

[1]*School of Physical Science and Technology, Soochow University, Soochow University, Suzhou 215006, China*
[2]*Institute of Electromagnetics and Acoustics and Key Laboratory of Electromagnetic Wave Science and Detection Technology, Xiamen University, Xiamen 361005, China*
[3]*College of Science, Nanjing University of Aeronautics and Astronautics, Nanjing 211106, China*
*yyfu@nuaa.edu.cn;
* leigao@suda.edu.cn;
* ydxu@suda.edu.cn

**Abstract:** Nonlinear metasurfaces offer new paradigm for boosting optical effect beyond limitations of conventional materials. In this work, we present an alternative way to produce pronounced third-harmonic generation (THG) based on nonlinear field resonances rather than linear field enhancement, which is a typical strategy for achieving strong nonlinear response. By designing and studying a nonlinear plasmonic-graphene metasurface at terahertz regime with hybrid guided modes and bound states in the continuum modes, it is found that a THG with a narrow bandwidth can be observed, thanks to the strong resonance between generated weak THG field and these modes. Such strong nonlinear field resonance greatly enhances the photon-photon interactions, thus resulting in a large effective nonlinear coefficient of the whole system. This finding provides new opportunity for studying nonlinear optical metasurfaces.

## 1. Introduction

Optical metasurfaces [1-4] have provided unprecedented capacities for arbitrary control of light behaviors [5-20], allowing a number of intriguing optical phenomena or metadevices, such as the generalized Snell's law [5], metalenses [6], the photonic spin Hall effect [8], wavefront controlling [10,11], and perfect anomalous diffraction [13-15]. In particular, metasurfaces with nonlinear responses [16,17] have attracted extensive attention, as their abilities offer a new paradigm to study nonlinear optics. Compared with conventional three-dimensional (3D) nonlinear bulk materials or metamaterials [17], two-dimensional (2D) metasurfaces can not only produce stronger nonlinear effect, but also do not need the complex phase-matching techniques that are usually required by the 3D counterparts. To obtain strong nonlinear response in a subwavelength scale, more intensive light-matter interaction need to be implemented in the metasurfaces with strong resonances [18-24], such as Fano resonances [18,19], bound states in the continuum (BIC) [21] and anapole resonances in high-index dielectric particles [23,24]. Owing to relatively high conversion efficiency [17], nonlinear metasurfaces are very promising in practical applications.

Third-order harmonic generation (THG), as a typical example of frequency conversion, is one of the most fundamental manifestations of nonlinear optical response, and offers opportunities for many applications in photonics, material science, all-optical signal processors and biosensing. A variety of nonlinear metasurfaces [25-29] have been demonstrated theoretically and experimentally to enhance the THG, including plasmonic metasurfaces, all-dielectric metasurfaces and dielectric-metal hybrid structures. In an optical nonlinear system, including metasurfaces, the linear and nonlinear process happen at the same time, and they are in no particular order. The linear process, such as reflection, refraction or scattering, is the result of photon-electron interactions, while the nonlinear process is due to the photon-photon interactions that are intrinsically weak and can only be realized at high intensity of light.

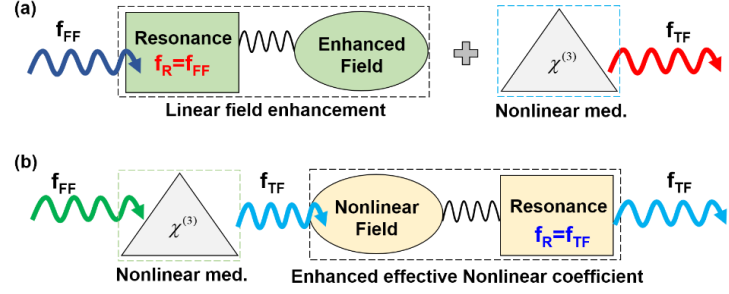

**Fig. 1**. Concepts of two different physical processes for nonlinear response. (a) Typical way. An incident light with FF coincides to the specific resonances (for example BICs) of designed linear optical system, i.e., $f_{FF} = f_R$, leading to a strong localization and enhancement of local electric field. The interaction of such strong electric field with the nonlinear medium produces pronounced THG with $f_{TF} = 3f_R$. (b) Another way proposed in this work. A FF light is incident into an optical system integrated with nonlinear medium, directly triggering a weak THG field with $f_{TF} = 3f_{FF}$. When the THG coincides to a specific resonance of designed optical system, i.e., $f_{TF} = f_R$, the interaction of such produced THG and the specific resonances and the associated feedback between linear and nonlinear processes lead to enhanced effective nonlinear coefficient.

So far, most of nonlinear metasurfaces for boosted THG response rely on the enhancement of linear local electric field (i.e., photon-electron interactions), and its typical physical process is schematically shown in Fig. 1(a). A linear optical system with a specific resonance mode is well designed at an operating frequency $f_R$. When a fundamental frequency (FF) light with $f_{FF} = f_R$ impinges upon it, the linear local field surrounding the nonlinear medium is greatly enhanced by strong resonances ($\vec{E}_l^R$). The interaction of such strong local field and the nonlinear medium leads to boosted THG, i.e., $\vec{P} \sim \chi^{(3)}(\vec{E}_l^R \cdot \vec{E}_l^R)\vec{E}_l^R$, with its third harmonic frequency $f_{TH} = 3f_{FF}$. For instance, by utilizing spoof surface plasmon resonances operated in a hybrid graphene-plasmonic grating at terahertz (THz) frequency [28], one can observe a perfect absorption and associated local field enhancement at $f_{FF} = 8.8$ THz. Accordingly, a pronounced THG at $f_{TH} = 3f_{FF} = 26.4$ THz is produced by the enhanced linear field along the graphene monolayer with nonlinear conductivity. Similarly, it has been shown that a nonlinear metasurface governed by BIC mode [29] can enable extremely large local field. Once the BIC mode is excited, a strong THG with $f_{TH} \approx 3f_R$ can be observed at $f_{FF} \approx f_R$.

In this work, we design and study a plasmonic-graphene nonlinear metasurface, which can exhibit strong THG response not in this typical way. We contribute such strong THG response to nonlinear coefficient amplification originating from nonlinear field resonance. Figure 1(b) schematically illustrates its concept. Likewise, in order to obtain a strong interaction between light and matter, a linear metasurface system with a specific resonant frequency $f_R$ is designed. Distinguished from the above-mentioned typical way, the frequency of incident FF light is $f_{FF} = f_R / 3$ instead of $f_{FF} = f_R$, far from the specific resonance. Due to the added nonlinear medium, the incident FF light with $f_{FF} = f_R / 3$ will produce a weak THG with $f_{TH} = 3f_{FF} = f_R$, which implies that the third harmonic frequency exactly coincides with the resonance frequency of the system. As a result, the interaction of such produced THG and the specific resonances leads to boosted THG. This seemingly simple operation is actually highly nontrivial because the THG field resonance greatly enhances the photon-photon interactions and effectively amplify the nonlinear coefficient of the entire system, (i.e., $\chi_e^{(3)} > \chi^{(3)}$). The two ways for strong THG in Fig. 1 seems to be the result of a simple interchange of the order of linear and nonlinear processes, while they differ fundamentally in physics. We will show that the THG based on our discussed way in some cases may be more efficient than that based on the linear local field enhancement corresponding to the typical way.

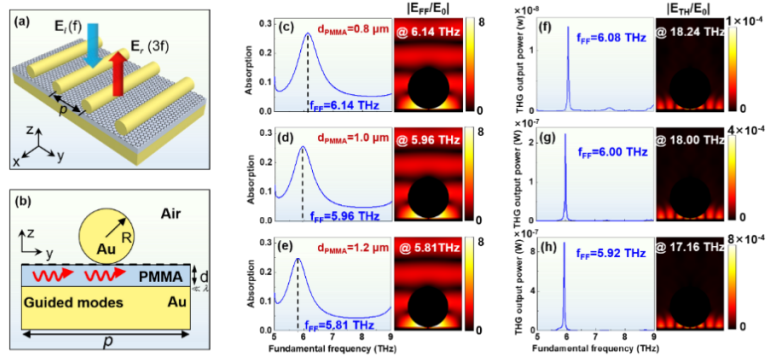

**Fig. 2.** (a) Schematic diagram of the considered nonlinear optical metasurface with its unit cell in (b). (c)-(e) are the linear absorption spectra of the nonlinear optical metasurface for the PMMA with different thickness d (the left panels), and the right panels show the electric field patterns at the localized surface plasmon (LSP) resonances. (f)-(g) show the corresponding output power of the THG as a function of the FF, with the right panels illustrating the electric filed distributions of THG at the resonances. In all cases, $p = 60\,\mu m$, $R = 18\,\mu m$ and $E_F = 0.3\,eV$.

## 2. Results and discussions

Let us start with the considered nonlinear metasurface, as shown in Fig. 2(a). The designed metasurface is a periodic array of gold (Au) cylinder with a radius $R$, laid on a graphene/dielectric/metal substrate along the $y$ direction. As the designed metasurface is 2D structure, the cylinders are infinite along $x$ direction. Figure 2(b) shows the details of a unit cell with a period $p$. The optical property of Au at terahertz regime is described by $\varepsilon_m = \varepsilon_\infty - f_p^2/(f^2 - i\gamma f)$, where $\varepsilon_\infty = 1.53$, $f_p = 2069\,THz$ and $\gamma = 17.65\,THz$ [30]. The dielectric layer is deeply subwavelength ($d \ll \lambda$). For the sake of simplicity, it is assumed to be dispersionless and lossless within the considered frequency range, and its permittivity is $\varepsilon_d = 2.25$. The monolayer graphene with Fermi level $E_F$ serves as the nonlinear material, and the conductivity of graphene is expressed by: $\sigma_g = \sigma_0 + \sigma^{(3)}|E|^2$, where $\sigma_0$ and $\sigma^{(3)}$ are the linear term and the third-order nonlinear term, respectively [31]. For the linear term $\sigma_0$, when $\hbar\omega/E_F \leq 2$, the interband transition of electrons in graphene is forbidden and only the intraband transition is left to dominate its optical response with incident light. As a result, $\sigma_0 = (e^2 E_F / \pi\hbar^2)[i/(\omega + i\tau^{-1})]$, where $\omega = 2\pi f$ is the angular frequency, $e$ is the electron charge, $\hbar$ is the reduced Planck constant, and $\tau$ is the relaxation time which is estimated as $10^{-13}$ s [32]. For the nonlinear term $\sigma^{(3)}$, its exact expression is taken from Ref. [33]. A transverse-magnetic (TM) polarized FF light (i.e., the magnetic field only along the $x$ direction) is normally incident on this nonlinear metasurface from air. COMSOL Multiphysics is used to study the nonlinear effect of the proposed metasurface, and the nonlinear graphene is modeled by a nonlinear surface current with $J = \sigma_0 E_{TH} + \sigma^{(3)} E_{FF}^3$ [34], where $E_{FF}$ and $E_{TH}$ are the electric fields of the FF and the TH, respectively. Based on the simulation scheme used in Ref. [28], we can obtain THG radiation in COMSOL Multiphysics.

Intuitively, due to localized surface plasmons (LSPs) [35], the electromagnetic (EM) energy of incident light can be guided into the bottom of each Au cylinder, leading to a strong local field near it. Figures 2(c)-2(e) numerically show the linear absorption spectra of FF for $d = 0.8$,

1.0 and 1.2 $\mu m$, respectively, where the intensity of incident light is 10 KW/cm². As the strong nonlinear effect of graphene at THz regime is approximately located in the region from 1 to 10 THz, the parameters of $p = 60\,\mu m$ and $R = 18\,\mu m$ are numerically found to design the metasurface with specific resonances in this region. Clearly, owing to the LSP resonances (see the field pattern in the right panels), a resonance peak appears in each absorption spectrum (see the left panel). According to the typical way, this strong local field will significantly enhance the interaction between light and nearby nonlinear graphene, producing a strong THG. However, the understanding of THG enhancement effect found in our designed nonlinear metasurface is not so straightforward. Figures 2(f)-2(h) display the corresponding results of THG output power as a function of FF (see the left panels). Strong THG emerges at $f_{FF} = 6.08$, 6.00 and 5.92 THz for $d = 0.8$, 1.0 and 1.2 $\mu m$, respectively, which slightly deviate from the corresponding LSP resonances of $f_{FF} = 6.14$, 5.96 and 5.81 THz in Figs. 2(c)-2(e). The corresponding electric field patterns of THG are plotted in the right panels. Most energy of the THG is bounded at the surface of the graphene/PMMA, featured with a standing wave with nine nodes. Note that although the FF of THG resonance is very close to the LSP resonance frequency and the difference is very small, the THG peak in the nonlinear spectrum is not caused by the LSP resonance. We will show later that such frequency difference is ubiquitous, and can be altered further by changing the geometric parameters, such as the thickness of dielectric layer and the radius of the metal cylinders (discussed later in Fig. 5 and Fig. 6).

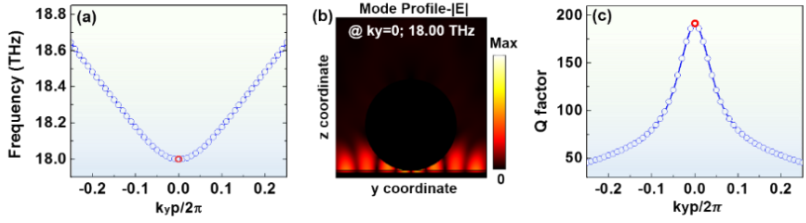

**Fig. 3.** Physical mechanism for enhanced THG based on hybrid guided mode. (a) Calculated band structure of the photonic structure in Fig. 2(a) around 18.00 THz. The blue circles indicate the hybrid guided modes in the considered photonic structure and they are out of the light line (i.e., in free space). (b) Electric-filed profile of the hybrid guided mode at $k_y$=0 with its corresponding eigenfrequency at 18.00 THz. (c) The Q factor of the band structure shown in (a). In all analysis, $p = 60\,\mu m$, $R = 18\,\mu m$, $d = 1.0\,\mu m$ and $E_F = 0.3\,eV$.

To uncover the mechanism of seemly unusual THG, the photonic band diagram of the considered structure was numerically calculated, as shown in Fig. 3(a). Here we take the case of $d = 1\,\mu m$ as an example for illustrations. In calculations, we only considered the optical properties of the purely linear system, thus removing the nonlinear term of graphene, as the nonlinear effect is weaker compared to the linear one. Owing to the interaction between the metal cylinder array and the dielectrics, a plasmonic-photonic band is plotted in Fig. 3(a), and we call it hybrid guided mode for convenience because these hybrid modes can propagate freely in the deep subwavelength dielectric layer along the $y$ direction. The eigenfrequency at $\Gamma$ point ($k_y = 0$) is about 18.0 THz and the corresponding eigenmode profile is shown in Fig. 3(b), which is exactly consistent with the field distribution of THG at $f_{FF} = 6.0\,THz$ in Fig. 2(g). In addition, we also calculated the $Q$ factor of this band, as shown in Fig. 3(c). The $Q$ factor at $k_y = 0$ is the largest, approximately $Q = 200$, which is almost the same as the $Q$ value calculated from THG spectrum in Fig. 2(g). Therefore, these results of photonic band diagram clearly reveal that the physical mechanism of the significant THG in the designed nonlinear metasurface is attributed to hybrid guided mode resonance of THG, rather than the enhancement of the linear local field. In physics, the THG effect depends on the term $\chi^{(3)}\vec{E}\cdot\vec{E}$, where $\vec{E}$ are linear local field, and $\chi^{(3)}$ is third order susceptibility. Note that the THG output power at $f_{FF}=5.96\,THz$ (corresponding to LSP resonances) and $f_{FF}=6.0$ THz, is $\approx 0.32\times 10^{-7}$ and $\approx 2.2\times 10^{-7}$ W (the corresponding conversion efficiency is about $3.7\times 10^{-11}$), respectively. In fact, the linear field enhancement in both cases is comparable. But in the case of $f_{FF}=6.0$ THz, the nonlinear field resonance stemming from hybrid guided mode resonance of THG greatly enhances the photon-photon interactions, thus resulting in an effective third order susceptibility greater than the intrinsic one of graphene, i.e., $\chi_{eff}^{(3)}>\chi^{(3)}$. This is the reason that the THG output power at $f_{FF}=6.0\,THz$ is larger than that at $f_{FF}=5.96\,THz$. Therefore, by designing mode resonance of third harmonic frequency in a linear system, the boosted THG could be obtained as well.

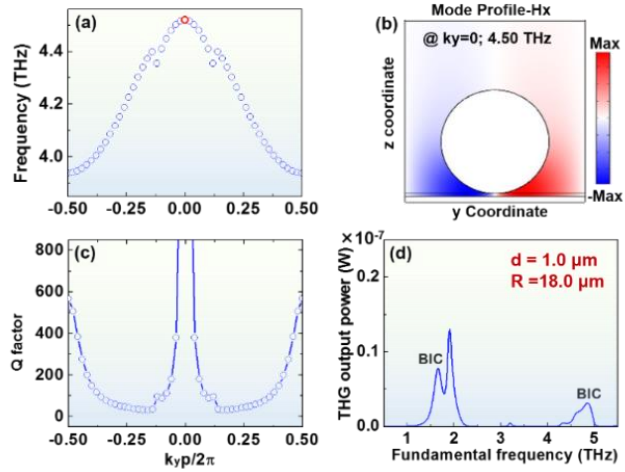

**Fig. 4**. Quasi-BIC based THG enhancement. (a) Calculated band structure of the photonic structure around 4.0 THz. (b) Electric-filed profile of quasi-BICs at $k_y=0$ with its corresponding eigenfrequency at 4.5 THz. (c) The Q factor of the band structure shown in (a). (d) Output power of the THG as a function of the FF. The two peaks labeled by BIC are at $f_{FF}\approx 1.6$ THz and $f_{FF}\approx 4.8$ THz. Here $p=60\,\mu m$, $R=18\,\mu m$, $d=1.0\,\mu m$ and $E_F=0.3\,eV$.

Alternatively, as our designed metasurface has mirror symmetry with respect to $z$ axis, the complete decoupling between incident wave and the eigenmode of odd symmetry can happen. As a result, the quasi-BIC mode or BIC mode can exist in our metasurface system, which can greatly boost THG due to their extremely strong resonances [29]. To find it, we numerically analyzed the photonic band structure of the considered structure. As shown in Fig. 4(a), it is a band of odd mode (see Fig. 4(b)), where the mode at $k_y=0$ indicated by a red circle is quasi-BIC mode with eignefrequency at about 4.5 THz. This point is confirmed by calculating the quality factor $Q$ of the whole band structure. As shown in the Fig. 4(c), the $Q$ factor reaches an extremely large value at $k_y=0$. Note that if the Ohmic's loss in metal is removed, i.e., $\gamma=0$, the obtained $Q$ value goes to infinity as $k_y\to 0$.

Now let us explore what happens to THG caused by BIC modes in the current case. Based on the two different ways for THG shown in Fig. 1, two resonant frequencies of $f_{FF}=1.5$ and 4.5 THz might be available to obtain enhanced THG via BIC modes. Figure 4(d) illustrates the calculated results of THG output power as a function of FF. Three THG resonances emerge in the spectrum; two resonances at $f_{FF}\approx 1.6$ and 4.8 THz are generated by the BIC mode and one resonance at $f_{FF} \approx 1.96$ THz is caused by the generated THG field resonance with above mentioned LSPs. Specifically, the peak around 1.6 THz, which is very close to $f_{TT} = 1.5$ THz (i.e., 4.5/3 THz), is caused by the resonance between THG field and BIC mode. While the peak around 4.8 THz, which is slightly off the eigenfrequency 4.5 THz of BIC, is caused by the local field enhancement resulting from the resonance between the incident light and BIC. Owing to the nature of BIC itself, a small deviation is seen in both cases, which is consistent with the results observed in [29], that is, the strongest nonlinear effect does not occur exactly at BIC point in the spectrum, but at a position near it. As we know, the $Q$ factor of BIC is infinite, which means that the incident EM energy cannot effectively couple to the optical system through BIC resonance. Strictly speaking, the energy coupled to the system is zero, which results in extremely weak THG. Therefore, in this sense, a true BIC is not an ideal choice for enhancing the interaction between light and matter. By contrast, the quasi-BICs with a moderate $Q$ factor are more favorable to enhance the interaction between light and matter. Moreover, one can see that in the current case, the THG output power at $f_{TT} \approx 1.6$ THz (the corresponding conversion efficiency is about $1.3 \times 10^{-12}$) is larger than that at $f_{TT} \approx 4.8$ THz. It implies that the THG based on our proposed way may be more efficient than that based on the typical way.

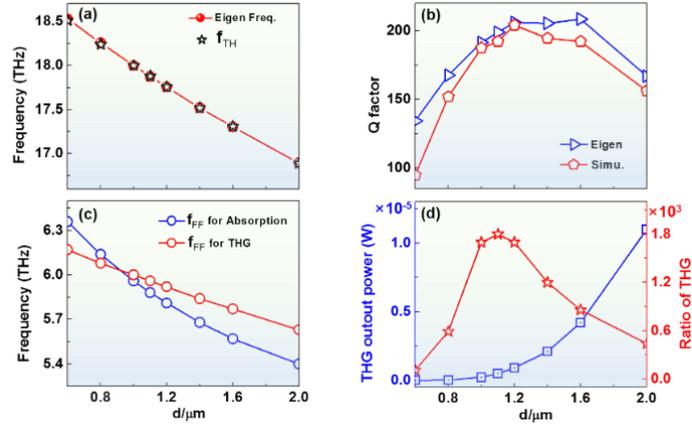

**Fig. 5**. Influence of dielectric layer thickness on THG. (a) The frequency of THG resonance and the eigenfrequency of hybrid guided mode; and (b) their Q factors. (c) FF of LSP-based absorption and the FF of THG resonance vs dielectric thickness . (d) The THG output power of the designed nonlinear metasurface and its ratio to that without the metal cylinder array. Here $p = 60 \, \mu m$, $R = 18 \, \mu m$, and $E_F = 0.3 \, eV$.

The THG resonance frequency can be adjusted by changing the thickness of the dielectric layer or the radius of the metal cylinders, because the hybrid guided modes and BIC modes depend on the geometry of the considered structure. Figure 5(a) numerically show the eigenfrequency (the red circles) of hybrid guided mode and the resonance frequency of THG (the black five stars) for the thickness $d$ ranging from 0.6 to 2.0 $\mu m$. It is clearly seen that as $d$ increases, both frequencies decrease monotonically, and both results exactly agree with each other, which further prove the validity of our proposed way for THG. Figure 5(b) shows their corresponding $Q$ factors, which undergo an increased and decreased process as $d$ increases. To further confirm that the enhanced THG does not stem from LSP resonances, we calculated the resonance frequency of LSP-based absorption for different thickness, which is displayed by the blue circles in Fig. 5(c). For comparison, the fundamental frequency (the red circles) corresponding to THG resonances in Fig. 5(a) are added. It is clearly seen that as $d$ changes, two curves follow different variation trends and only intersect at $d \approx 0.94 \, \mu m$. Moreover, to test the performance of THG enhancement, Fig. 5(d) shows the relationship between the THG output power and the thickness of the dielectric layer (the blue curve), as well as the ratio (the red curve) of this output power to THG output power of nonlinear metasurface without Au cylinder array. The maximum value of ratio is about $1.8 \times 10^3$, occurring at $d \approx 1.1\,\mu m$, which is mainly attributed to the quality of the hybrid guided modes. We also carried out a study on the influence of the radius of the metal cylinders. The calculated results are shown in Fig. 6, which are similar with those in Fig. 5. For instance, as $R$ changes, the eigenfrequency (the red circles) of hybrid guided mode and the resonance frequency (the five stars) of THG exactly agree with each other (see Fig. 6(a)). The FF of LSP-based absorption (the blue circles) and the FF corresponding to THG resonances (the red circles) also follow different variation trends, with the intersection point at $R \approx 17.6\,\mu m$. Likewise, for BIC modes, the calculated results (not shown here) demonstrate that the THG resonance frequency shifts with the change of geometric size.

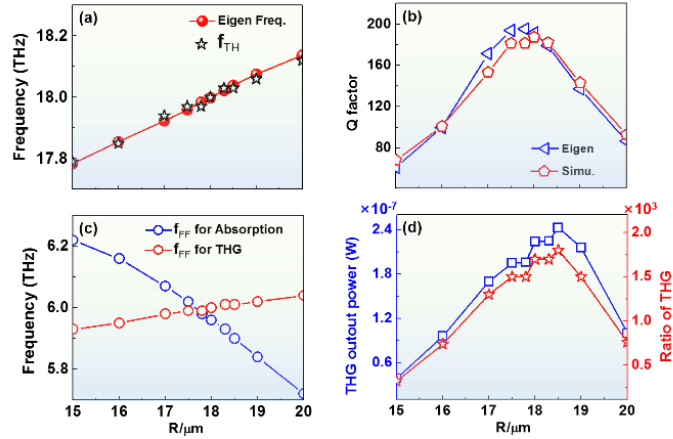

**Fig. 6**. Influence of Radius of metal cylinder on THG. (a) the frequency of THG resonance and the eigenfrequency of hybrid guided mode; and (b) their Q factor. (c) FF of LSPs-based absorption and the FF of THG resonance vs the radius of cylinder. (d) The THG output power of designed nonlinear metasurface and its ratio to that of designed nonlinear metasurface without the metal cylinder array. Here $p = 60\,\mu m$, $d = 1\,\mu m$, and $E_F = 0.3\,eV$.

## 3. Conclusion

In conclusion, we have examined and discussed a new way to obtain enhanced nonlinear response by designing and studying a plasmonic-graphene nonlinear metasurface at terahertz regime. We have shown that due to the existence of hybrid guided modes and quasi-BIC modes with resonance frequency $f_R$ in the designed structure, a significant THG can be observed at $f_{FF} = f_R/3$, and the maximum conversion efficiency can be enhanced by three orders. Different from the typical way of local field enhancement, the proposed mechanism for enhanced THG relies on strong resonance of the generated THG field with these specific modes, which greatly magnify the effective nonlinear coefficient of the whole system via photon-photon interactions. With respect to the conversion efficiency based on our proposed way, we think it should be discussed case by case, as the conversion efficiency is highly depending on the features of resonance modes and can be improved by carrying out the optimized studies. In addition, owing to the lower incident frequency, our proposed way makes the nonlinear elements further miniaturized, showing some advantages in the integrated application of photonic devices. Similar strategy can be extended to study other nonlinear effect, such as second-order harmonic generation.

**Funding**

National Natural Science Foundation of China (11974010, 11904169, 118743111, 11774252 and 1604229); the Natural Science Foundation of Jiangsu Province (BK20161210, BK20171206 and BK20190383); a project funded by the China Postdoctoral Science Foundation (grant no. 2018T110540); the Qing Lan project; the "333" project (BRA2015353).

**Acknowledgments**

Y. Xu thanks the support from the state key laboratory of matamaterial electromagnetic modulation technology (grant no. GYL08-1458), and the Priority Academic Program Development (PAPD) of Jiangsu Higher Education Institutions.